\documentclass[letter, twocolumn]{jpsj3}

\usepackage[dvipdfmx]{graphicx}
\usepackage{float}
\usepackage{amsmath, amssymb}
\usepackage{fancybox}
\usepackage{ascmac}
\usepackage{physics}
\usepackage{txfonts}
\usepackage{color}
\usepackage{bm}

\newcommand{\Eq}[1]{Eq.~(\ref{#1})}

\newcommand{\Fig}[1]{Fig.~\ref{#1}}

\title{Nonlinear nonreciprocal electronic conductivity driven by magnetic field gradients}
\author{Taisei Yamanaka$^1$, Yoshihiko Ihara$^2$, and Satoru Hayami$^2$}
\inst{
  $^1$Department of Physics, Hokkaido University, Sapporo 060-0810, Japan\\
  $^2$Graduate School of Science, Hokkaido University, Sapporo 060-0810, Japan }
\abst{ 
We theoretically propose the emergence of nonlinear nonreciprocal conductivity in centrosymmetric paramagnetic systems when a spatially gradient magnetic field is externally applied. 
The key essence lies in the appearance of magnetic toroidal dipole moment under the gradient field that breaks both spatial inversion and time-reversal symmetries. 
By analyzing a minimal tight-binding model on a two-dimensional system, we show that an effective coupling between the 
magnetic toroidal dipole moment arising from the gradient field and sublattice-dependent antisymmetric spin--orbit interaction plays an important role in inducing the nonlinear nonreciprocal transport.
We also discuss the favorable situation to observe the nonlinear nonreciprocal conductivity in real materials by presenting an experimental setup in order to stimulate the findings.
}

\begin{document}
\maketitle

The interplay among charge, spin, and orbital degrees of freedom in electrons gives rise to diverse physical phenomena in condensed matter physics. 
Depending on the types of symmetry broken, corresponding electronic ordered states appear and drive various physical phenomena. 
Although the appearance of physical phenomena is accounted for by the breakings of the crystalline and time-reversal symmetries under the point group according to Neumann's principle, it is often difficult to extract the microscopic essence beyond the symmetry when the internal electronic degrees of freedom are strongly entangled. 

The multipole representation theory has been developed in order to understand the relationship between the microscopic electronic degrees of freedom and macroscopic physical phenomena. 
Such a representation theory has been originally developed for the $f$-electron system with the spatial inversion symmetry~\cite{Santini_RevModPhys.81.807, kuramoto2009multipole, Kusunose_JPSJ.77.064710}.
The concept of multipoles has been extended to more general cases so as to include the odd-parity hybridization and cluster degrees of freedom in the symmetry-adapted form~\cite{Suzuki_PhysRevB.95.094406, suzuki2018first, hayami2018microscopic, Hayami_PhysRevLett.122.147602, kusunose2020complete, Kusunose_PhysRevB.107.195118}. 
In the multipole representation, there are four types of multipoles with distinct spatial inversion and time-reversal symmetries: electric, magnetic, magnetic toroidal (MT), and electric toroidal multipoles, which constitute a complete basis set in arbitrary physical space~\cite{kusunose2022generalization, hayami2024unified}. 
The multipole representation provides a deep understanding of unconventional physical phenomena in materials, such as the anomalous Hall/Nernst effect by the magnetic octupole ordering in Mn$_3$Sn~\cite{nakatsuji2015large, ikhlas2017large, kuroda2017evidence, higo2018anomalous, higo2018large, nakatsuji2022topological} and Mn$_3$$A$N ($A =$  Ga, Sn, and Ni)~\cite{Gurung_PhysRevMaterials.3.044409, Zhou_PhysRevB.99.104428, Boldrin_PhysRevMaterials.3.094409, Huyen_PhysRevB.100.094426, you2021cluster}, the metallic-type magneto-electric effect and nonlinear transverse conductivity by the MT dipole ordering in UNi$_4$B~\cite{saito2018evidence, ota2022zero} and Ce$_3$TiBi$_5$~\cite{shinozaki2020magnetoelectric, shinozaki2020study}, and the Edelstein effect by the electric toroidal monopole in tellurium~\cite{yoda2015current, furukawa2017observation, yoda2018orbital, Suzuki_PhysRevB.107.115305}.

The responses against external stimuli are important to examine the role of the microscopic electronic degrees of freedom in materials. 
Along this line, various external fields have been used, such as the electric, magnetic, and strain fields, which couple to the electric dipole, magnetic dipole, and electric quadrupole degrees of freedom, respectively. 
Meanwhile, it is difficult to investigate the nature of higher-rank and toroidal-type multipole moments by static external fields, since these moments do not directly couple to the spatially uniform external fields. 
It is desired to investigate the nature of unconventional multipoles by using external fields with the spatial gradient. 

\begin{figure}[!t]
  \centering
  \includegraphics[width = \linewidth]{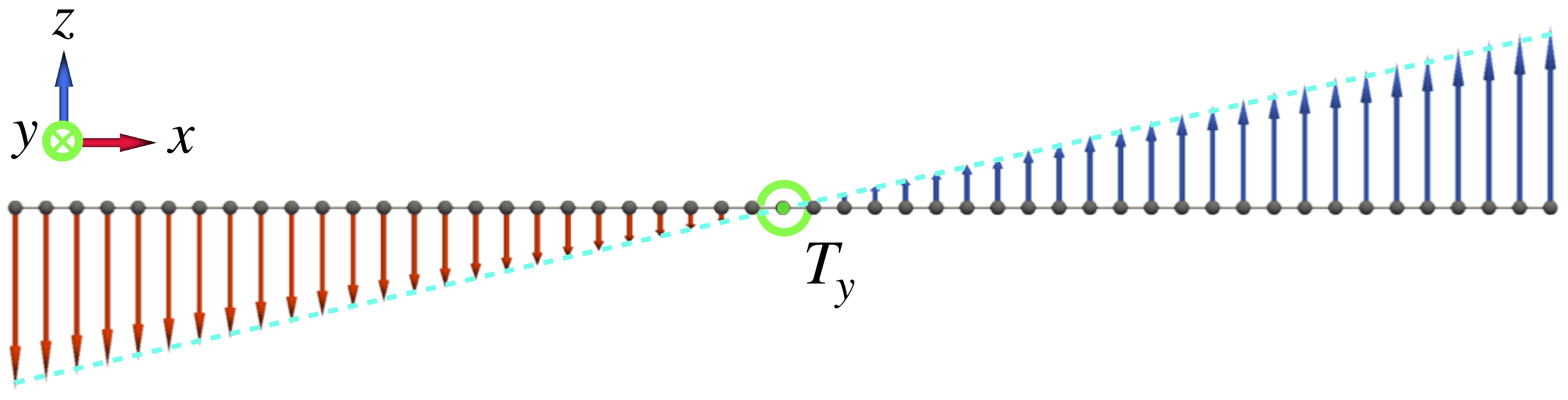}
  \caption{(Color online)
  Schematic picture of the magnetic toroidal dipole moment $T_y$ induced by the magnetic field gradient $\partial B_z/\partial x$. The blue and red arrows represent the positive and negative $z$ spin moments, respectively. 
  }
  \label{fig:model1}
\end{figure}

In this Letter, we theoretically propose that a magnetic field gradient is a good platform to investigate the nature of the MT dipole moment~\cite{schmid2001ferrotoroidics, kopaev2009toroidal, Spaldin_0953-8984-20-43-434203, cheong2018broken, hayami2024unified, xu2024magnetic, Azimi-Mousolou_PhysRevB.110.L140403}. 
The MT dipole can couple to the magnetic field gradient by its time-reversal-odd and polar-vector nature. 
According to the expression for the MT dipole, $ \bm{T} \propto \bm{r}_i \times \bm{S}_i$ ($\bm{r}_i$ and $\bm{S}_i$ represent the position and spin vectors at site $i$, respectively), its $y$ component is induced in the presence of the magnetic field gradient given by $ \partial B_z/\partial x$, as schematically shown in Fig.~\ref{fig:model1}. 
We demonstrate the emergent MT dipole under the magnetic field gradient by analyzing the minimal tight-binding model consisting of the dimer-chain system. 
We show that the band dispersion is asymmetrically modulated along the MT dipole moment direction under the magnetic field gradient when both field and gradient directions are perpendicular to the MT dipole moment direction
,and the emergent MT dipole moment induces the nonlinear nonreciprocal transport. 
Moreover, we find that the sublattice-dependent antisymmetric spin-orbit interaction is important to induce the nonlinear nonreciprocal transport. 
Finally, we discuss an experimental setup to realize the theoretical proposal in order to stimulate the findings. 

\begin{figure}[!t]
  \centering
  \includegraphics[width=\linewidth]{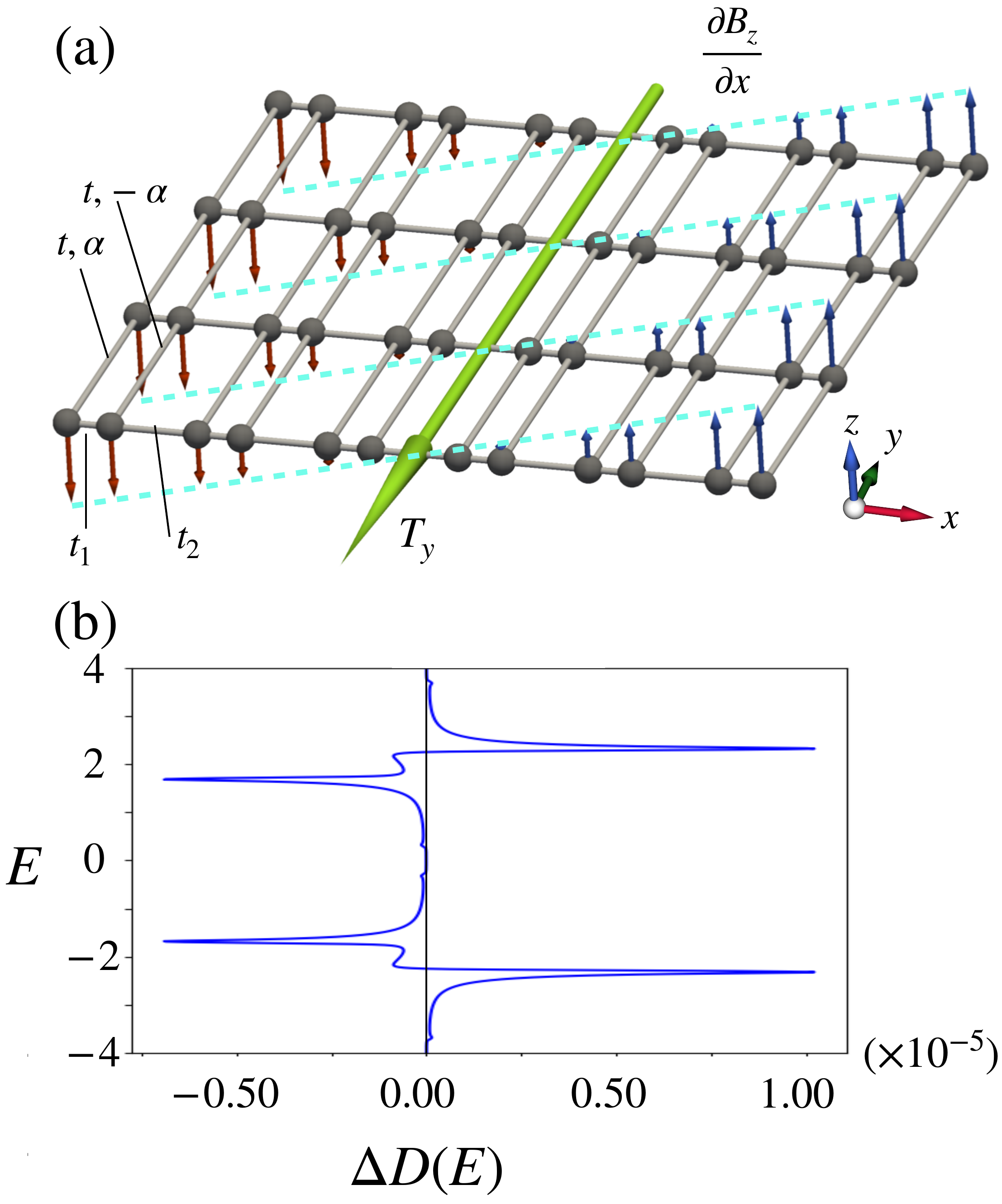}
  \caption{(Color online)
  (a) Two-dimensional dimer-chain system under the magnetic field gradient, leading to the MT dipole along the $y$ direction, $T_y$.
  The hopping parameters in the model in \Eq{eq:Hamiltonian} are also shown.
  (b) 
  The difference in density of states between the positive wave-number contribution $D_{+}(E)$ and the negative wave-number contribution $D_{-}(E)$ at $\Delta 
  \tilde{B} = 2\times10^{-5}$, which is represented by $ \Delta D(E) = D_{+}(E) - D_{-}(E)$.
  }
  \label{fig:model2}
\end{figure}

Let us consider the following tight-binding model in the two-dimensional dimer-chain system. 
As shown in Fig.~\ref{fig:model2}, we consider the effect of the magnetic field gradient along the $x$ direction; accordingly, we adopt the open (periodic) boundary conditions for the $x$ ($y$) direction; we set the number of the lattice site along the $x$ direction to $N$ and the lattice constant along the $y$ direction to unity.  
Then, the model Hamiltonian is given by
\begin{align}
  \label{eq:Hamiltonian}
  \mathcal{H} &= \mathcal{H}_{\mathrm{kin}} + \mathcal{H}_\mathrm{ASOI} + \mathcal{H}_\mathrm{Z}, \\
  \label{eq: Hkin}
  \mathcal{H}_\mathrm{kin} &= \sum_{i=1}^{N} \sum_{k\sigma}^{}
   \left[ \varepsilon(k){c^\dag_{i, k\sigma}} c_{i, k\sigma} - t_{ij}(c^\dag_{i, k\sigma} c_{j, k\sigma} + \mathrm{H.c.})\right], \\
  \begin{split}
  \mathcal{H}_\mathrm{ASOI} &=  
  \sum_{i=1}^{N/2} \sum_{k\sigma\sigma'}^{} \vb*{g}(k)\cdot \vb*{\sigma}^{\sigma\sigma'} 
 \left({c^\dag_{2i-1, k\sigma}} c_{2i-1, k\sigma^{'}} - c^\dag _{{2i, k\sigma} }c_{{2i, k\sigma^{'}} }\right),
  \end{split}\\
  \label{eq: Hamzeeman}
  \mathcal{H}_\mathrm{Z} &= - \sum_{i=1}^{N} \sum_{k}^{} \tilde{B}_{i}  
  (c^\dag_{i, k\uparrow} c_{i, k\uparrow}-c^\dag_{i, k\downarrow} c_{i, k\downarrow}), 
\end{align}
where $c^\dag_{i, k\sigma} (c_{i, k\sigma})$ is the creation (annihilation) operator for site $i$ along the $x$ direction, wave number $k_y \equiv k$ along the $y$ direction, and spin $\sigma = \uparrow, \downarrow$.
The kinetic Hamiltonian $\mathcal{H}_{\mathrm{kin}}$ in Eq.~(\ref{eq: Hkin}) consists of the inter-chain hopping along the $y$ direction in the first term and the intra-chain hopping along the $x$ direction in the second term; $\varepsilon(k)$ is given by $\varepsilon(k) = -2t\cos k$ and $t_{ij}$ includes the intra-dimer hopping $t_1$ ($j=i\pm 1$) and the inter-dimer hopping $t_2$ ($j=i\pm2$), as shown in Fig.~\ref{fig:model2}(a).
We choose $t_1=1$ as the energy unit of the model, and set $t_2=0.7$ and $t=1$.

The second term $\mathcal{H}_\mathrm{ASOI}$ in Eq.~(\ref{eq:Hamiltonian}) represents the antisymmetric spin-orbit interaction (ASOI) that arises from the synergy between the relativistic spin-orbit coupling and the local crystalline electric field (CEF) at the lattice site~\cite{yanase2014magneto,hayami2014spontaneous,hayami2016emergent}; $\vb*{g}(k)$ is referred to as the $g$-vector and is generally related to the local CEF as $\vb*{k} \times (\bm{\nabla} V) \sim \vb*{g}(\vb*{k})$, where $\bm{k}$ is a three-dimensional wave vector and $\bm{\nabla} V$ is the potential gradient. 
$g$-vector satisfies $\bm{g}(-k)=-\bm{g}(k)$. 
Since the dimer-chain structure leads to $\bm{\nabla}V \parallel \hat{\bm{x}}$, the $g$-vector is represented by $\vb*{g}(k) = (0, 0, \alpha \sin k)$ in the two-dimensional system, where $\alpha$ is the magnitude of the ASOI.

The third term $\mathcal{H}_\mathrm{Z}$ in Eq.~(\ref{eq:Hamiltonian}) represents the effect of the magnetic field gradient with $\tilde{B}_i=\mu_{\rm B}B$; $\mu_{\rm B}$ is the Bohr magneton and we set the $g$-factor for the electron to $2$. 
We define $\tilde{B}_i = (-N/2+i-1/2)\Delta \tilde{B}$ for $1\leq i \leq N$ to include the spatially dependent Zeeman coupling. 
$\Delta \tilde{B}$ represents the difference of the magnetic field between the neighboring sites and $\bm{\sigma}$ represents the vector of the Pauli matrix in spin space. 
It is noted that $\mathcal{H}_\mathrm{Z}$ is invariant under the $\mathcal{PT}$ symmetry, where $\mathcal{P}$ and $\mathcal{T}$ represent the spatial inversion and time-reversal operations, respectively.  
In the following, we consider $N=3000$ and take the grid points along the $y$ direction (the number of $k$ points) as $N_k=4000$.

When the magnetic field gradient is turned on, both the $\mathcal{P}$ and $\mathcal{T}$ symmetries are broken while keeping the $\mathcal{PT}$ symmetry, which indicates the emergence of the MT dipole along the $y$ direction. 
Accordingly, the energy band structure is modulated so as to be asymmetric in terms of the $k_y$ direction~\cite{yanase2014magneto, Hayami_doi:10.7566/JPSJ.84.064717}. 
To demonstrate that, we show the difference between the density of states in terms of the positive $k$ component, $D_+ (E)$, and the negative $k$ component, $D_- (E)$, i.e., $\Delta D (E)= D_{+}(E) - D_{-}(E)$, in Fig.~\ref{fig:model2}(b). 
Note that $\Delta D (E)=0$ when the band structure is symmetric in terms of $k$. 
$\Delta D (E)$ becomes finite when the magnetic field gradients have a finite value ($\Delta \tilde{B} = 2\times10^{-5}$ in this study) and the MT dipole appears. 
Thus, the nonlinear nonreciprocal transport is expected along the $y$ direction~\cite{wakatsuki2017nonreciprocal, tokura2018nonreciprocal, watanabe2020nonlinear, yatsushiro2022analysis, Suzuki_PhysRevB.105.075201, nagaosa2024nonreciprocal}. 

In the presence of the MT dipole with the breakings of both $\mathcal{P}$ and $\mathcal{T}$ symmetries, electric conductivity has the finite second-order contribution, which is defined by $ j_\mu = \sigma_{\mu\nu\eta} E_\nu E_\eta$; $j_\mu$ and $E_{\nu}$ are an electric current and an electric field for the $\mu, \nu, \eta = x, y, z$ direction, respectively. 
We evaluate the nonlinear longitudinal conductivity $\sigma_{yyy}$ by using the Boltzmann equation with the relaxation time approximation~\cite{yasuda2016large}, which is given by 
\begin{align}
  \label{sigma}
  \begin{split}
  \sigma_{yyy}
   &= \left(\frac{e \tau}{\hbar}\right)^2 \frac{1}{N N_{k}} \sum_{n,k} \langle n,k | \hat{j}_{y} | k,n \rangle  \frac{\partial^2 f[\epsilon_n(k)]}{\partial k^2},\\
  \end{split}
\end{align}
where $e$, $\tau$, and $\hbar$ are the electron charge, relaxation time, and the reduced Planck constant, respectively; we take $e = \tau = \hbar = 1$. 
$f[\epsilon_n({k})]$ is the Fermi distribution function and $| k,n \rangle$ is the eigenstate with the wave number $k$ and the band index $n$. $\hat{j}_{y}$ is the current operator given by $-\frac{e}{\hbar}\frac{\partial \mathcal{H}}{\partial k}$.

\begin{figure}[t!]
  \centering
  \includegraphics[width=\linewidth]{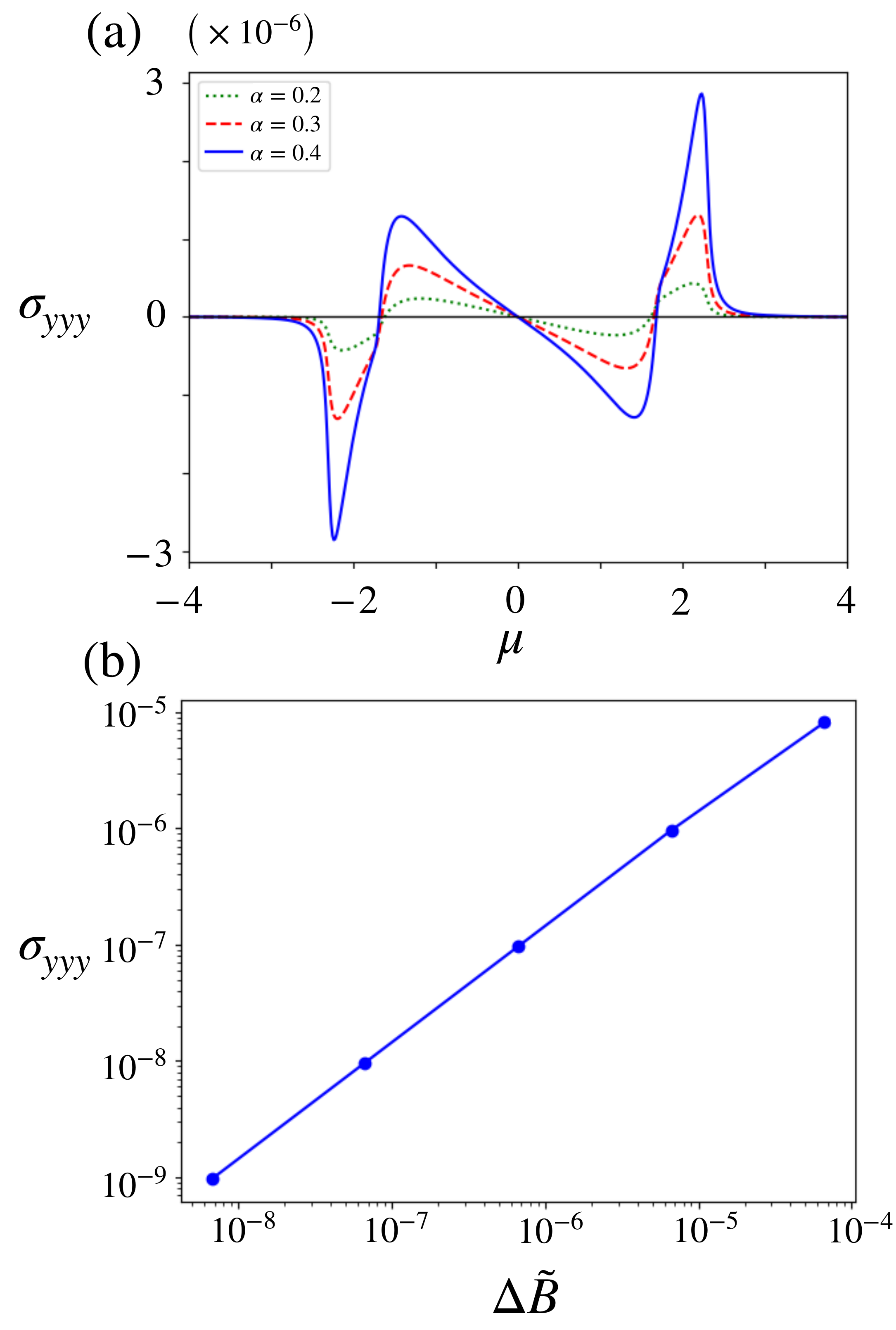}
  \caption{(Color online)
  (a) $\mu$ dependence of $\sigma_{yyy}$ at $\alpha = 0.2, 0.3, $ and $ 0.4 $ at $\Delta \tilde{B} = 2\times10^{-5}$. 
  (b) $\Delta \tilde{B}$ dependence of $\sigma_{yyy}$ at $\alpha = 0.4$ and $\mu = 2.24$.
  }
  \label{fig:cond_dep}
\end{figure}

Figure~\ref{fig:cond_dep}(a) shows the $\mu$ dependence of $\sigma_{yyy}$ for $\alpha=0.2$, $0.3$, and $0.4$ at $\Delta \tilde{B} = 2 \times 10^{-5}$. 
The results show that the nonlinear nonreciprocal conductivity becomes finite for $\Delta \tilde{B} \neq 0$. 
Moreover, we find that $\sigma_{yyy}$ becomes larger as the ASOI $\alpha$ increases, which indicates the importance of the ASOI in inducing $\sigma_{yyy}$. 
Indeed, $\sigma_{yyy}$ vanishes for $\alpha=0$, since the asymmetric band deformation does not occur in this case. 
In addition, large $\sigma_{yyy}$ is obtained when the asymmetricity in the band structure is large, as compared to the result in Fig.~\ref{fig:cond_dep}(a) with that in Fig.~\ref{fig:model2}(b); their peak positions are located at almost the same energy. 
In this way, the magnetic field gradient induces the nonlinear nonreciprocal conductivity through ASOI and associated asymmetric band deformation. 

Next, we show the $\Delta \tilde{B}$ dependence of $\sigma_{yyy}$ at $\alpha=0.4$ and $\mu=2.24$ in Fig.~\ref{fig:cond_dep}(b).
The data indicates that $\sigma_{yyy}$ linearly increases as $\Delta \tilde{B}$ increases over four orders of magnitudes, which suggests that the larger magnetic field gradient is favorable to experimentally detect the signal of the nonlinear transport.

\begin{figure}[h]
  \centering
  \includegraphics[width=\linewidth]{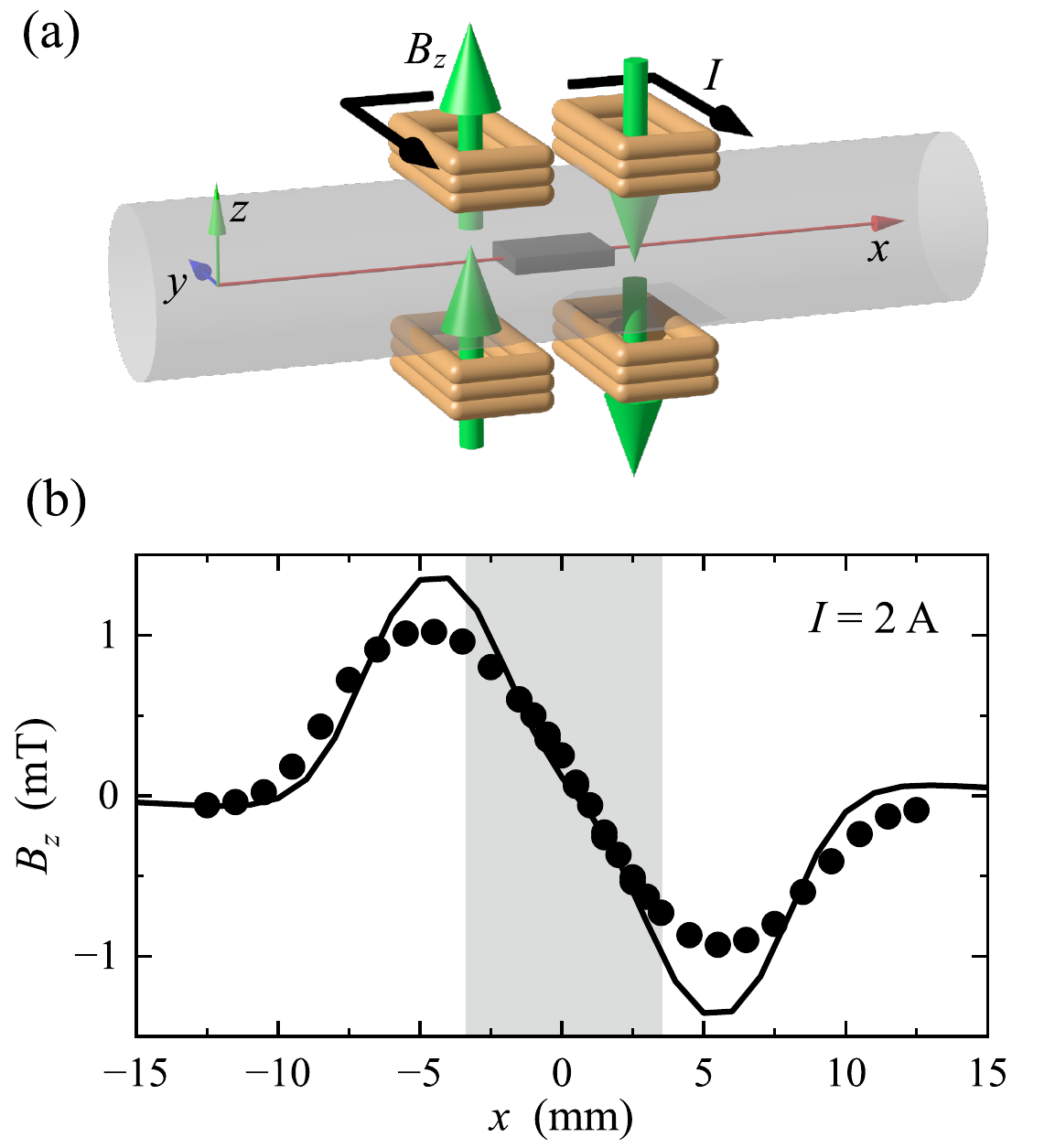}
  \caption{(Color online)
    (a) A design for the gradient coil. The electric current flows toward the opposite directions for the coils places side by side and generate the gradient field at the sample position. 
    (b) Spatial distribution of the magnetic field along $z$ direction measured on the $x$ axis shown in (a). Solid line and black points are the result of simulation and measurement by Hall sensor, respectively. The coil current $I$ was set to 2 A. 
  }
  \label{fig:experiment}
\end{figure}

Finally, let us present an experimental setup for applying the desired magnetic field gradient to a realistic material. 
Figure~\ref{fig:experiment}(a) illustrates a gradient coil that generates the gradient magnetic field being $\partial B_z/ \partial x \simeq $ constant around the sample position.  
Two square coils wound in opposite directions are placed above and below the sample space. 
The magnetic field along $z$ direction on the $x$ axis is calculated and shown in Fig.~\ref{fig:experiment}(b) by a solid line. 
The gradient coil is actually prepared and the field distribution under the coil current $I=2$ A is measured along the $x$ axis by a Hall sensor (black points in Fig.~\ref{fig:experiment}).  
The field gradient within the shadowed region ($x<\pm 3$ mm) is approximately 0.3 T/m.  
Although the magnitude of the magnetic field gradient, which is in the order of $10^{-9}$~T per atom, is much smaller than that used in the model, it would be improved up to $10^{-6}$~T per atom by increasing coil current and reducing sample space.

\begin{figure}[t!]
  \centering
  \includegraphics[width=\linewidth]{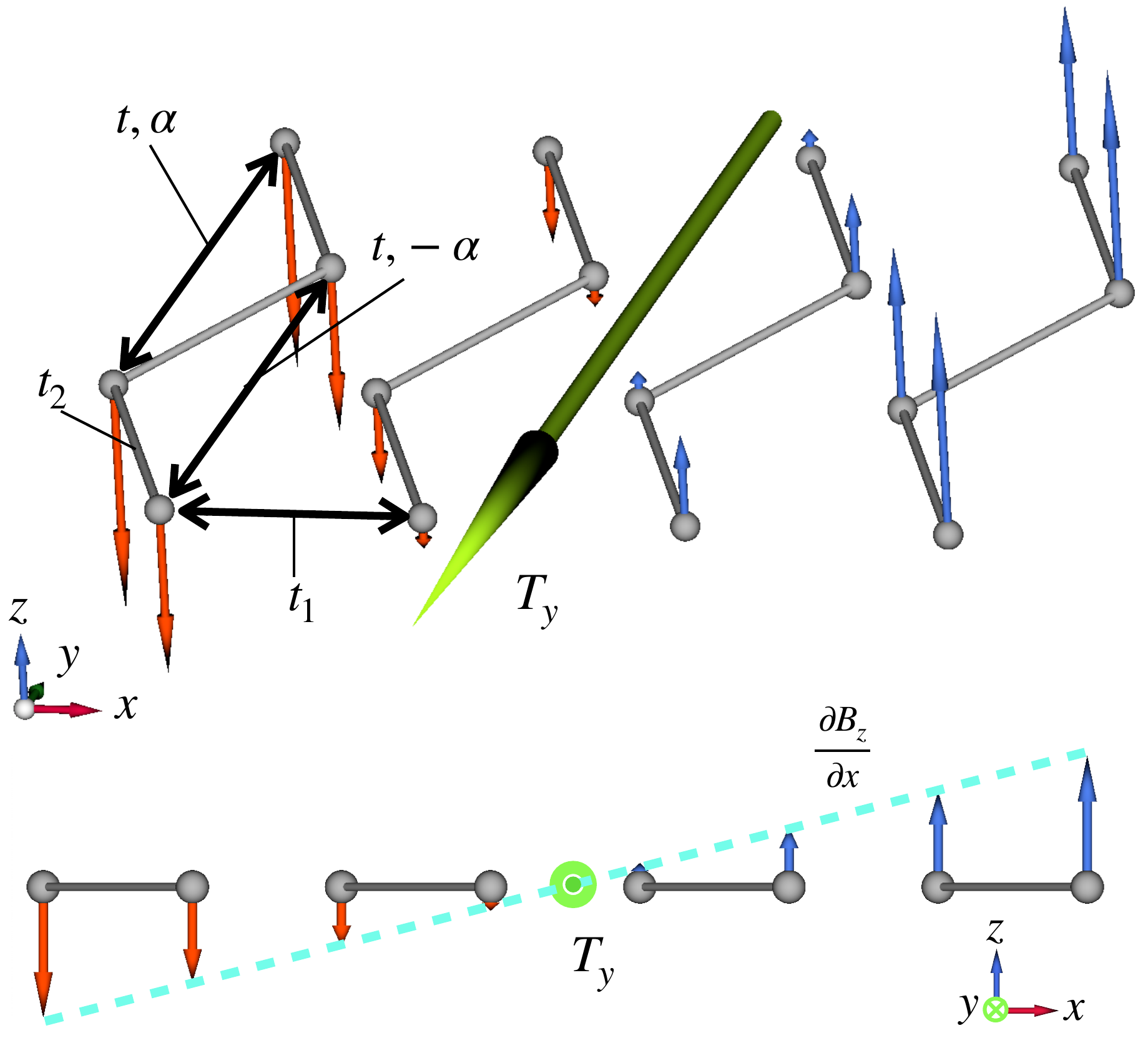}
  \caption{(Color online)
 (Top panel) Two-dimensional zigzag chain on which the model in Fig.~\ref{fig:model2}(a) is mapped. 
 (Bottom panel) Side view in (a). 
  }
  \label{fig:model_zigzag}
\end{figure}

The results in Figs.~\ref{fig:cond_dep}(a) and \ref{fig:cond_dep}(b) suggest that two important conditions are posed to the target material to observe the sizable nonlinear signal in addition to the large slope of the gradient field. 
The one is the large ASOI and the other is the small bandwidth so as to effectively increase the effect of magnetic field. 
The sublattice-dependent ASOI would be found in a system, where the spacial inversion symmetry is locally broken, which includes the zigzag chain~\cite{yanase2014magneto, Hayami_doi:10.7566/JPSJ.85.053705, cysne2021orbital}, honeycomb~\cite{li2013coupling, Yanagi_PhysRevB.97.020404, Matsumoto_PhysRevB.101.224419}, and diamond~\cite{Hayami_PhysRevB.97.024414, Ishitobi_doi:10.7566/JPSJ.88.063708, Winkler_PhysRevB.107.155201} structures. 
In particular, the present model is the best mapped onto the zigzag chain, as shown in \Fig{fig:model_zigzag}, where the same model parameters $(t, t_1, t_2, \alpha)$ can be used by appropriately replacing the wave-number dependence.
Thus, the zigzag-chain materials, such as NdRu$_2$Al$_{10}$ and TbRu$_2$Al$_{10}$~\cite{thiede1998ternary, Tanida_PhysRevB.84.115128, ishii2012successive, yanase2014magneto, mizushima2015metamagnetic}, are the best candidates for experimental realization. 
The observation of the nonlinear nonreciprocal transport based on the present mechanism will be left as an intriguing issue in the future study.

In summary, we have investigated the effect of the magnetic field gradient to a material with ASOI on the basis of the microscopic model analysis.
Our result suggests that the magnetic field gradient, which conjugates with the MT dipole moment, introduces the MT-dipole-related physical phenomena, namely the nonlinear nonreciprocal conductivity. 
To demonstrate that, we have analyzed the minimal tight-binding model in the dimer-chain system. 
By calculating the second-order electrical conductivity within the framework of the Boltzmann equation with the relaxation time approximation, we found that the ASOI arising from the local inversion symmetry breaking plays an important role. 
The nonlinear nonreciprocal response increases with the magnetic field gradient, demanding huge gradient fields, for which an actual design for the gradient coil is presented. 
The present results indicate the possibility of inducing and controlling the MT dipole moment by the static magnetic field, which will stimulate further studies related to the MT-dipole physics, such as the linear magnetoelectric effect~\cite{popov1999magnetic, EdererPhysRevB.76.214404, thole2018magnetoelectric}, nonlinear nonreciprocal transport~\cite{Sawada_PhysRevLett.95.237402, Kawaguchi_PhysRevB.94.235148, Watanabe_PhysRevX.11.011001, Hayami_PhysRevB.106.014420}, and nonlinear (spin) Hall effect~\cite{Wang_PhysRevLett.127.277201, Shao_PhysRevLett.124.067203, Liu_PhysRevLett.127.277202, hayami2022nonlinear, Kondo_PhysRevResearch.4.013186}. 

\begin{acknowledgments} 
This research was supported by JSPS KAKENHI Grants Numbers JP21H01037, JP22H00101, JP22H01183, JP23H04869, JP23K03288, JP23K20827, and by JST CREST (JPMJCR23O4) and JST FOREST (JPMJFR2366). 
\end{acknowledgments}

\bibliographystyle{jpsj}
\bibliography{main}

\begin{thebibliography}{10}

\bibitem{Santini_RevModPhys.81.807}
P.~Santini, S.~Carretta, G.~Amoretti, R.~Caciuffo, N.~Magnani, and G.~H. Lander: Rev. Mod. Phys. {\bfseries 81} (2009) 807.

\bibitem{kuramoto2009multipole}
Y.~Kuramoto, H.~Kusunose, and A.~Kiss: J. Phys. Soc. Jpn. {\bfseries 78} (2009) 072001.

\bibitem{Kusunose_JPSJ.77.064710}
H.~Kusunose: J. Phys. Soc. Jpn. {\bfseries 77} (2008) 064710.

\bibitem{Suzuki_PhysRevB.95.094406}
M.-T. Suzuki, T.~Koretsune, M.~Ochi, and R.~Arita: Phys. Rev. B {\bfseries 95} (2017) 094406.

\bibitem{suzuki2018first}
M.-T. Suzuki, H.~Ikeda, and P.~M. Oppeneer: J. Phys. Soc. Jpn. {\bfseries 87} (2018) 041008.

\bibitem{hayami2018microscopic}
S.~Hayami and H.~Kusunose: J. Phys. Soc. Jpn. {\bfseries 87} (2018) 033709.

\bibitem{Hayami_PhysRevLett.122.147602}
S.~Hayami, Y.~Yanagi, H.~Kusunose, and Y.~Motome: Phys. Rev. Lett. {\bfseries 122} (2019) 147602.

\bibitem{kusunose2020complete}
H.~Kusunose, R.~Oiwa, and S.~Hayami: J. Phys. Soc. Jpn. {\bfseries 89} (2020) 104704.

\bibitem{Kusunose_PhysRevB.107.195118}
H.~Kusunose, R.~Oiwa, and S.~Hayami: Phys. Rev. B {\bfseries 107} (2023) 195118.

\bibitem{kusunose2022generalization}
H.~Kusunose and S.~Hayami: J. Phys.: Condens. Matter {\bfseries 34} (2022) 464002.

\bibitem{hayami2024unified}
S.~Hayami and H.~Kusunose: J. Phys. Soc. Jpn. {\bfseries 93} (2024) 072001.

\bibitem{nakatsuji2015large}
S.~Nakatsuji, N.~Kiyohara, and T.~Higo: Nature {\bfseries 527} (2015) 212.

\bibitem{ikhlas2017large}
M.~Ikhlas, T.~Tomita, T.~Koretsune, M.-T. Suzuki, D.~Nishio-Hamane, R.~Arita, Y.~Otani, and S.~Nakatsuji: Nat. Phys. {\bfseries 13} (2017) 1085.

\bibitem{kuroda2017evidence}
K.~Kuroda, T.~Tomita, M.-T. Suzuki, C.~Bareille, A.~Nugroho, P.~Goswami, M.~Ochi, M.~Ikhlas, M.~Nakayama, S.~Akebi, R.~Noguchi, R.~Ishii, N.~Inami, K.~Ono, H.~Kumigashira, A.~Varykhalov, T.~Muro, T.~Koretsune, R.~Arita, S.~Shin, T.~Kondo, and S.~Nakatsuji: Nat. Mater. {\bfseries 16} (2017) 1090.

\bibitem{higo2018anomalous}
T.~Higo, D.~Qu, Y.~Li, C.~Chien, Y.~Otani, and S.~Nakatsuji: Appl. Phys. Lett. {\bfseries 113} (2018) 202402.

\bibitem{higo2018large}
T.~Higo, H.~Man, D.~B. Gopman, L.~Wu, T.~Koretsune, O.~M. van't Erve, Y.~P. Kabanov, D.~Rees, Y.~Li, M.-T. Suzuki, S.~Patankar, M.~Ikhlas, C.~L. Chien, R.~Arita, R.~D. Shull, J.~Orenstein, and S.~Nakatsuji: Nat. Photonics {\bfseries 12} (2018) 73.

\bibitem{nakatsuji2022topological}
S.~Nakatsuji and R.~Arita: Ann. Rev. Condens. Matter Phys. {\bfseries 13} (2022) 119.

\bibitem{Gurung_PhysRevMaterials.3.044409}
G.~Gurung, D.-F. Shao, T.~R. Paudel, and E.~Y. Tsymbal: Phys. Rev. Mater. {\bfseries 3} (2019) 044409.

\bibitem{Zhou_PhysRevB.99.104428}
X.~Zhou, J.-P. Hanke, W.~Feng, F.~Li, G.-Y. Guo, Y.~Yao, S.~Bl\"ugel, and Y.~Mokrousov: Phys. Rev. B {\bfseries 99} (2019) 104428.

\bibitem{Boldrin_PhysRevMaterials.3.094409}
D.~Boldrin, I.~Samathrakis, J.~Zemen, A.~Mihai, B.~Zou, F.~Johnson, B.~D. Esser, D.~W. McComb, P.~K. Petrov, H.~Zhang, and L.~F. Cohen: Phys. Rev. Mater. {\bfseries 3} (2019) 094409.

\bibitem{Huyen_PhysRevB.100.094426}
V.~T.~N. Huyen, M.-T. Suzuki, K.~Yamauchi, and T.~Oguchi: Phys. Rev. B {\bfseries 100} (2019) 094426.

\bibitem{you2021cluster}
Y.~You, H.~Bai, X.~Feng, X.~Fan, L.~Han, X.~Zhou, Y.~Zhou, R.~Zhang, T.~Chen, F.~Pan, et~al.: Nat. Commun. {\bfseries 12} (2021) 6524.

\bibitem{saito2018evidence}
H.~Saito, K.~Uenishi, N.~Miura, C.~Tabata, H.~Hidaka, T.~Yanagisawa, and H.~Amitsuka: J. Phys. Soc. Jpn. {\bfseries 87} (2018) 033702.

\bibitem{ota2022zero}
K.~Ota, M.~Shimozawa, T.~Muroya, T.~Miyamoto, S.~Hosoi, A.~Nakamura, Y.~Homma, F.~Honda, D.~Aoki, and K.~Izawa: arXiv:2205.05555  (2022).

\bibitem{shinozaki2020magnetoelectric}
M.~Shinozaki, G.~Motoyama, M.~Tsubouchi, M.~Sezaki, J.~Gouchi, S.~Nishigori, T.~Mutou, A.~Yamaguchi, K.~Fujiwara, K.~Miyoshi, and Y.~Uwatoko: J. Phys. Soc. Jpn. {\bfseries 89} (2020) 033703.

\bibitem{shinozaki2020study}
M.~Shinozaki, G.~Motoyama, T.~Mutou, S.~Nishigori, A.~Yamaguchi, K.~Fujiwara, K.~Miyoshi, and A.~Sumiyama: JPS Conf. Proc. {\bfseries 30} (2020) 011189.

\bibitem{yoda2015current}
T.~Yoda, T.~Yokoyama, and S.~Murakami: Sci. Rep. {\bfseries 5} (2015) 12024.

\bibitem{furukawa2017observation}
T.~Furukawa, Y.~Shimokawa, K.~Kobayashi, and T.~Itou: Nat. Commun. {\bfseries 8} (2017) 954.

\bibitem{yoda2018orbital}
T.~Yoda, T.~Yokoyama, and S.~Murakami: Nano Lett. {\bfseries 18} (2018) 916.

\bibitem{Suzuki_PhysRevB.107.115305}
Y.~Suzuki and Y.~Kato: Phys. Rev. B {\bfseries 107} (2023) 115305.

\bibitem{schmid2001ferrotoroidics}
H.~Schmid: Ferroelectrics {\bfseries 252} (2001) 41.

\bibitem{kopaev2009toroidal}
Y.~V. Kopaev: Physics-Uspekhi {\bfseries 52} (2009) 1111.

\bibitem{Spaldin_0953-8984-20-43-434203}
N.~A. Spaldin, M.~Fiebig, and M.~Mostovoy: J. Phys.: Condens. Matter {\bfseries 20} (2008) 434203.

\bibitem{cheong2018broken}
S.-W. Cheong, D.~Talbayev, V.~Kiryukhin, and A.~Saxena: npj Quantum Mater. {\bfseries 3} (2018) 19.

\bibitem{xu2024magnetic}
X.~Xu, F.-T. Huang, and S.-W. Cheong: J. Phys.: Condens. Matter {\bfseries 36} (2024) 203002.

\bibitem{Azimi-Mousolou_PhysRevB.110.L140403}
V.~Azimi-Mousolou, A.~Bergman, A.~Delin, O.~Eriksson, M.~Pereiro, D.~Thonig, and E.~Sj\"oqvist: Phys. Rev. B {\bfseries 110} (2024) L140403.

\bibitem{yanase2014magneto}
Y.~Yanase: J. Phys. Soc. Jpn. {\bfseries 83} (2014) 014703.

\bibitem{hayami2014spontaneous}
S.~Hayami, H.~Kusunose, and Y.~Motome: Phys. Rev. B {\bfseries 90} (2014) 081115.

\bibitem{hayami2016emergent}
S.~Hayami, H.~Kusunose, and Y.~Motome: J. Phys. : Cond. Mat. {\bfseries 28} (2016) 395601.

\bibitem{Hayami_doi:10.7566/JPSJ.84.064717}
S.~Hayami, H.~Kusunose, and Y.~Motome: J. Phys. Soc. Jpn. {\bfseries 84} (2015) 064717.

\bibitem{wakatsuki2017nonreciprocal}
R.~Wakatsuki, Y.~Saito, S.~Hoshino, Y.~M. Itahashi, T.~Ideue, M.~Ezawa, Y.~Iwasa, and N.~Nagaosa: Science advances {\bfseries 3} (2017) e1602390.

\bibitem{tokura2018nonreciprocal}
Y.~Tokura and N.~Nagaosa: Nat. Commun. {\bfseries 9} (2018) 3740.

\bibitem{watanabe2020nonlinear}
H.~Watanabe and Y.~Yanase: Phys. Rev. Res. {\bfseries 2} (2020) 043081.

\bibitem{yatsushiro2022analysis}
M.~Yatsushiro, R.~Oiwa, H.~Kusunose, and S.~Hayami: Phys. Rev. B {\bfseries 105} (2022) 155157.

\bibitem{Suzuki_PhysRevB.105.075201}
Y.~Suzuki: Phys. Rev. B {\bfseries 105} (2022) 075201.

\bibitem{nagaosa2024nonreciprocal}
N.~Nagaosa and Y.~Yanase: Annual Review of Condensed Matter Physics {\bfseries 15} (2024) 63.

\bibitem{yasuda2016large}
K.~Yasuda, A.~Tsukazaki, R.~Yoshimi, K.~Takahashi, M.~Kawasaki, and Y.~Tokura: Phys. Rev. Lett. {\bfseries 117} (2016) 127202.

\bibitem{Hayami_doi:10.7566/JPSJ.85.053705}
S.~Hayami, H.~Kusunose, and Y.~Motome: J. Phys. Soc. Jpn. {\bfseries 85} (2016) 053705.

\bibitem{cysne2021orbital}
T.~P. Cysne, F.~S.~M. Guimar\~aes, L.~M. Canonico, T.~G. Rappoport, and R.~B. Muniz: Phys. Rev. B {\bfseries 104} (2021) 165403.

\bibitem{li2013coupling}
X.~Li, T.~Cao, Q.~Niu, J.~Shi, and J.~Feng: Proc. Natl. Acad. Sci. U.S.A. {\bfseries 110} (2013) 3738.

\bibitem{Yanagi_PhysRevB.97.020404}
Y.~Yanagi, S.~Hayami, and H.~Kusunose: Phys. Rev. B {\bfseries 97} (2018) 020404.

\bibitem{Matsumoto_PhysRevB.101.224419}
T.~Matsumoto and S.~Hayami: Phys. Rev. B {\bfseries 101} (2020) 224419.

\bibitem{Hayami_PhysRevB.97.024414}
S.~Hayami, H.~Kusunose, and Y.~Motome: Phys. Rev. B {\bfseries 97} (2018) 024414.

\bibitem{Ishitobi_doi:10.7566/JPSJ.88.063708}
T.~Ishitobi and K.~Hattori: J. Phys. Soc. Jpn. {\bfseries 88} (2019) 063708.

\bibitem{Winkler_PhysRevB.107.155201}
R.~Winkler and U.~Z\"ulicke: Phys. Rev. B {\bfseries 107} (2023) 155201.

\bibitem{thiede1998ternary}
V.~M. Thiede, T.~Ebel, and W.~Jeitschko: J. Mater. Chem. {\bfseries 8} (1998) 125.

\bibitem{Tanida_PhysRevB.84.115128}
H.~Tanida, D.~Tanaka, M.~Sera, S.~Tanimoto, T.~Nishioka, M.~Matsumura, M.~Ogawa, C.~Moriyoshi, Y.~Kuroiwa, J.~E. Kim, N.~Tsuji, and M.~Takata: Phys. Rev. B {\bfseries 84} (2011) 115128.

\bibitem{ishii2012successive}
I.~Ishii, Y.~Suetomi, H.~Muneshige, S.~Kamikawa, T.~K.~Fujita, S.~Tanimoto, T.~Nishioka, and T.~Suzuki: J. Phys. Soc. Jpn. {\bfseries 81} (2012) 064602.

\bibitem{mizushima2015metamagnetic}
T.~Mizushima, Y.~Watanabe, J.-i. Ejiri, T.~Kuwai, and Y.~Isikawa: J. Phys.: Conf. Ser., Vol. 592, 2015, p. 012051.

\bibitem{popov1999magnetic}
Y.~F. Popov, A.~Kadomtseva, D.~Belov, G.~Vorob'ev, and A.~Zvezdin: J. Exp. Theor. Phys. Lett. {\bfseries 69} (1999) 330.

\bibitem{EdererPhysRevB.76.214404}
C.~Ederer and N.~A. Spaldin: Phys. Rev. B {\bfseries 76} (2007) 214404.

\bibitem{thole2018magnetoelectric}
F.~Th{\"o}le and N.~A. Spaldin: Philos. Trans. R. Soc. A {\bfseries 376} (2018) 20170450.

\bibitem{Sawada_PhysRevLett.95.237402}
K.~Sawada and N.~Nagaosa: Phys. Rev. Lett. {\bfseries 95} (2005) 237402.

\bibitem{Kawaguchi_PhysRevB.94.235148}
H.~Kawaguchi and G.~Tatara: Phys. Rev. B {\bfseries 94} (2016) 235148.

\bibitem{Watanabe_PhysRevX.11.011001}
H.~Watanabe and Y.~Yanase: Phys. Rev. X {\bfseries 11} (2021) 011001.

\bibitem{Hayami_PhysRevB.106.014420}
S.~Hayami and M.~Yatsushiro: Phys. Rev. B {\bfseries 106} (2022) 014420.

\bibitem{Wang_PhysRevLett.127.277201}
C.~Wang, Y.~Gao, and D.~Xiao: Phys. Rev. Lett. {\bfseries 127} (2021) 277201.

\bibitem{Shao_PhysRevLett.124.067203}
D.-F. Shao, S.-H. Zhang, G.~Gurung, W.~Yang, and E.~Y. Tsymbal: Phys. Rev. Lett. {\bfseries 124} (2020) 067203.

\bibitem{Liu_PhysRevLett.127.277202}
H.~Liu, J.~Zhao, Y.-X. Huang, W.~Wu, X.-L. Sheng, C.~Xiao, and S.~A. Yang: Phys. Rev. Lett. {\bfseries 127} (2021) 277202.

\bibitem{hayami2022nonlinear}
S.~Hayami, M.~Yatsushiro, and H.~Kusunose: Phys. Rev. B {\bfseries 106} (2022) 024405.

\bibitem{Kondo_PhysRevResearch.4.013186}
H.~Kondo and Y.~Akagi: Phys. Rev. Research {\bfseries 4} (2022) 013186.

\end{thebibliography}

\end{document}